\documentclass[aip,pop,sd,reprint]{revtex4-1}

\usepackage{graphicx}
\usepackage{bm}
\usepackage{color}
\usepackage{amsmath}

\newcommand{\nc}{\newcommand}
\nc{\rcite}[1]{Ref.~\cite{#1}}
\nc{\rcites}[1]{Refs.~\cite{#1}}
\nc{\eqeqref}[1]{Eq.~\eqref{eq:#1}}
\nc{\eqseqref}[2]{Eqs.~\eqref{eq:#1}-\eqref{eq:#2} }
\nc{\secref}[1]{Sec.~\ref{sec:#1}}
\nc{\secsref}[2]{Sec.~\ref{sec:#1}-Sec.~\ref{sec:#2}}
\nc{\ssecref}[1]{Sec.~\ref{ssec:#1}}
\nc{\ssecsref}[2]{Sec.~\ref{ssec:#1}-Sec.~\ref{ssec:#2}}

\draft 

\begin{document}


\title{Plasma turbulence in the equatorial electrojet:  A two-dimensional Hamiltonian fluid model} 



\author{Ehab Hassan}
\email[correspondence email: ]{ehab@sci.asu.edu.eg}
\affiliation{Physics, Ain Shams University, EG}
\author{I. Keramidas Charidakos}
\affiliation{University of Colorado at Boulder, CO, US}
\author{P.J. Morrison}
\affiliation{Physics Department, University of Texas at Austin, TX, US}
\affiliation{Institute for Fusion Studies, University of Texas at Austin, TX, US}
\author{D.R. Hatch}
\author{W. Horton}
\affiliation{Institute for Fusion Studies, University of Texas at Austin, TX, US}

\date{\today}

\begin{abstract}
 A nonlinear unified fluid model that describes the Equatorial Electrojet, including the Farley-Buneman and gradient-drift plasma instabilities, is defined and shown to be a  noncanonical Hamiltonian system.  Two geometric constants of motion for the model are obtained and shown to be  Casimir invariants. A reformulation of the model  shows the roles of the density-gradient scale-length ($L_n$) and the cross-field drift-velocity ($\upsilon_E$) in controlling the dynamics of  unstable modes in the growing, transition, and saturation phases of a simulation.
\end{abstract}


\maketitle 

\section{Introduction}\label{sec:Introduction}
The weakly ionized plasma of  the equatorial electrojet is characterized by two types of instabilities, the  Farley-Buneman and gradient-drift instabilities. The presence of these instabilities in the E-region makes the electrojet rich with plasma structures extending from kilometer to sub-meter scales.\cite{fejer1980ionospheric} The spectral and spatial characteristics of these instabilities have been identified using radar observations and \textit{in-situ} sounding rocket measurements. \cite{farley2009equatorial,fejer1980ionospheric,kudeki1987condor} 

In this paper, we identify the system Hamiltonian and associated Casimir invariants for a unified fluid model that captures the dynamics of both the Farely-Buneman and gradient-drift instabilities. This model has been described in Hassan \textit{et al.}\ in Ref.~\onlinecite{hassan2015multiscale} and shown to capture the most important properties observed in electrojet plasma fluctuations. In Ref.~\onlinecite{hassan2016multiscale} the authors proves that the unified fluid model conserves energy, and they found the role of the nonlinear terms in transferring energy to small structures of scale-length that can not be excited linearly.

The behavior of plasma in the equatorial electrojet can be modeled by the time evolution of the ion density and momentum, subject to the quasineutrality condition. One usually derives such dynamical equations by taking moments of a distribution function. The equations thus derived are fairly general since they contain the physical description of phenomena that occur over vastly different length and time scales. To reduce the complexity, the exact moment equations can be subsequently manipulated, according to the particular phenomenon one wishes to model, in order to filter out irrelevant dynamics in time and length scales. This process commonly takes the form of small parameter expansions and assumptions about the geometry of the system under consideration. Unfortunately, there is no rigorous prescription for this procedure and one has only his or her intuition to rely on. As a result, the systems of equations produced by such ad hoc procedures often come with a host of shortcomings. A very serious one is the potential loss of the Hamiltonian character: The parent model, that is, the system of charged particles interacting with an electromagnetic field, is Hamiltonian and as a consequence, it is desirable that any reduced description of it should retain this property.  The issue is not just a harmless question of mathematical formalism: The process of reduction might have introduced unwanted dissipation and as a result, the system might violate energy conservation at the ideal limit. By ideal limit, we refer to what remains from the system once all dissipative and source terms such as collisions, fluid models of Landau damping, and boundary terms have been discarded. To the contrary, a Hamiltonian system is guaranteed to conserve energy for closed boundary conditions. Indeed, as we can see in Ref.~\onlinecite{kimura2014energy, franck2015energy} out of all the different versions of implemented extended MHD models only some conserve energy whereas a survey of hybrid kinetic-MHD models in Ref.~\onlinecite{tronci2014hybrid} indicates that the use of the pressure coupling scheme results in a non-Hamiltonian model that not only fails to conserve energy but also contains a spurious, high frequency, Alfven wave instability. Similarly, inadequacies of gyrofluid models that do not conserve energy were noted in Ref.~\onlinecite{scott05}.

Nonetheless, energy conservation is not the sole reason one might have to pursue the discovery of the Hamiltonian formulation of a system. Hamiltonian theory, provides us with a set of tools which we can use to reveal important aspects of the system. One of the most crucial is the existence of geometric invariants known as Casimir invariants. Those are an artifact of the degeneracy of the cosymplectic operator and one can use them to construct free energy functionals whose minimization produces non-trivial equilibrium states. The existence of such states is not guaranteed in the absence of a Poisson bracket. In studying the dynamical system we are not using any particular methodology to derive the model in a manner that preserves its Hamiltonian character. Instead, we are simply lucky that the system is found to be Hamiltonian and we merely recover its bracket and cast it in its Hamiltonian form.

This paper is organized as follows: In \secref{PlasmaDynamics} we present the system geometry in the slab model \ssecref{geometry}, the set of nonlinear partial differential equations that govern the system's dynamics \ssecref{dynamicEquations}, and simulation results of the unified fluid model \ssecref{simulationResults}. The derivation of the system's Hamiltonian is shown in \secref{Hamiltonian}. In \secref{NonCanonicalHamiltonian} we prove that the dynamical equations form a noncanonical system by finding its Lie-Poisson bracket \ssecref{HamiltonianBracket} and showing that this bracket satisfies the Jacobi identity. The system Casimirs \ssecref{FindingCasimirs} and the reformulation of the system's dynamical equations in terms of a new variable \ssecref{newVariables} are shown in \secref{Casimirs}. Finally, 
we summarize the paper and draw our conclusions in \secref{conclusions}.

\section{The electrojet model}\label{sec:PlasmaDynamics}
\subsection{Geometry and background}\label{ssec:geometry}
The large electrical conductivity in the equatorial electrojet is attributable to the presence of a mixture of unmagnetized collisional ions and magnetized electrons \cite{baker1953electric}. The bounding of the E-region in the vertical direction by two layers of very small conductivity makes slab geometry a suitable treatment for modeling the equatorial electrojet \cite{farley1973instabilities,farley2009equatorial}. In this geometry, the plasma dynamics are studied in a plane transverse to the geomagnetic field that aligns with the x-direction, where the positive $x$-axis points northward. The plasma dynamical  plane is defined by the y-axis and z-axis that are pointing westward and upward, respectively. 

In the E-region, the isothermal state, $T_e=T_i$, and the quasineutrality condition, $n_e=n_i$, for plasma ions and electrons can be assumed.\cite{kelley2009earth} The dominance of $NO^+$ in the E-region over other ion species allows the consideration of the dynamics of a single ion in the plasma.\cite{schunk2009ionospheres} The ion  mean-free path is small because the ion-neutral collision frequency  ($\nu_{in}$) is large compared to the ion  gyrofrequency ($\omega_{ci}$). Therefore, the Lorentz force, $\bm{\upsilon}_i\times \bm{B}$, in the ions equation of motion can be ignored without any loss of generality, and the ion velocity is given by $\bm{\upsilon}_i=-{\bm\nabla}\chi$ as was done in Ref.~\onlinecite{kelley2009earth}.  On the other hand, the electrons are considered magnetized for the large ratio between their gyrofrequency ($\omega_{ce}$) and collision-frequency ($\nu_{en}$) with the neutral background \cite{kelley2009earth}.
\subsection{Equations of motion}\label{ssec:dynamicEquations}
A model was proposed by Hassan \textit{et al.}\ in Ref.~\onlinecite{hassan2015multiscale} to unify Farley-Buneman and gradient-drift instabilities in the equatorial electrojet. In that model, the plasma dynamics in the ion viscosity-tensor and electron polarization drifts were considered to play an important role in stabilizing the evolving fields. These two terms cause strong stabilization of the unstable linear growing modes of sub-meter scale-length \cite{hassan2015multiscale}, and excitation of the active small structures in the equatorial electrojet \cite{hassan2016multiscale}. As a result, the plasma dynamics in the linear regime give rise to growth-rate and phase velocity profiles that are comparable to those derived from the kinetic treatment of Farley-Buneman instability of Ref.~\onlinecite{schmidt1973density}. Also, the results in the saturation region in the nonlinear regime show good agreement with the radar and rocket observations. 

For electrostatic plasma waves, the ions continuity and momentum equations can be written as:
\begin{eqnarray}\label{eq:Uni3fIonContinuity}
\partial_tn&=&\bm{\nabla}\cdot\left(n\bm{\nabla}\chi\right)\,,
\\
\partial_t\chi&=&\upsilon_{t_i}^2\ln(n)+\frac{\Omega_{ci}}{B}\phi-\nu_{in}\chi
\nonumber
\\
&&\hspace{1cm} +\frac{1}{2}|{\bm\nabla}\chi|^2+\frac{4}{3}\frac{\upsilon_{t_i}^2}{\nu_{in}}\nabla^2\chi\,,
\label{eq:Uni3fIonMomentum}
\end{eqnarray}
where $\upsilon_{t_i}$ is the ion  thermal speed and the electrostatic field is given by ${\bf{E}}=-{\bm\nabla}\phi$.

The magnetized electron drift velocity, in a plane perpendicular to the geomagnetic field, can be given after some algebraic manipulation of electrons' equation of motion to have the following form \cite{hassan2015multiscale}:
\begin{equation}\label{eq:Uni3fElectronDrift}
{\bm\upsilon_{e\perp}}={\bm\upsilon_{E}}+{\bm\upsilon_{d_e}}+{\bm\upsilon_{P_e}}+{\bm\upsilon_{\nu}}\,,
\end{equation}
where, $\bm\upsilon_{E}$ is the ${\bf{E}}\times{\bf{B}}$ drift velocity, ${\bm\upsilon_{d_e}}$ is the diamagnetic drift velocity, ${\bm\upsilon_{P_e}}$ is the polarization drift velocity, and ${\bm\upsilon_{\nu}}$ is the drift velocity due to the frictional force between the electrons and the neutral background. 

Thus, using the ions and electrons drift velocities in the plasma quasineutrality condition, $\bm{\nabla}\cdot(\bm{J}_i 
+\bm{J}_e)=0$, the third dynamical equation is found in the following form:
\begin{flalign}\label{eq:Uni3fQuasineutrality}
\partial_t\nabla^2\phi&=\frac{T_e\nu_{en}}{e}\nabla^2\ln(n)-\nu_{en}\nabla^2\phi-B\Omega_{ce}\nabla^2\chi\nonumber\\
                      &-\Omega_{ce}\left[\phi,\ln(n)\right]-\frac{1}{B}\left[\phi,\nabla^2\phi\right]\nonumber\\
                      &+\frac{T_e\nu_{en}}{e}{\bm\nabla}\ln(n)\cdot{\bm\nabla}\ln(n)-\nu_{en}{\bm\nabla}\ln(n)\cdot{\bm\nabla}\phi\nonumber\\
                      &-B\Omega_{ce}{\bm\nabla}\ln(n)\cdot{\bm\nabla}\chi\,,
\end{flalign}
where, $\left[f,g\right]$ is the usual Poisson  bracket defined by $\left[f,g\right]=\partial_x{f}\partial_y{g}-\partial_x{g}\partial_y{f}$.

The set of partial differential \eqseqref{Uni3fIonContinuity}{Uni3fQuasineutrality} governs the plasma dynamics in the equatorial electrojet region that extends between 103 and 108 km in altitude. It also unifies the physics of the gradient-drift and Farley-Buneman instabilities which can be excited in this region depending on the solar and geophysical conditions.

\section{Dynamical System Hamiltonian}\label{sec:Hamiltonian}
In the dynamical plasma system of the unified fluid model for the Equatorial Electrojet instabilities the energy comes into the system from the top and bottom boundaries due to the non-zero gradients of the background density ($L_n^{-1} = \partial_z\ln n_o$) and electric potential ($\upsilon_E=-B_o^{-1}\partial_z\phi_o$) in the vertical direction which are considered constant energy sources. On the other hand, the energy is dissipated by collisions of electrons and ions with the background neutral particles.\cite{hassan2016multiscale}

To separate the sources of free energies in the dynamical system from the system's Hamiltonian and dissipation terms, we split the constant background density and electric potential from their fluctuating components. Thus we can rewrite the dynamical
\eqseqref{Uni3fIonContinuity}{Uni3fQuasineutrality} in the following form:
\begin{align}\label{eq:IonContinuitySeparate}
\begin{split}
\partial_t{\delta{n}} = \bm\nabla\cdot(\delta{n}\bm\nabla{\delta\chi}) + \bm\nabla\cdot(n_o\bm\nabla{\delta\chi})\,,
\end{split}					
\end{align}
\begin{align}\label{eq:IonMomentumSeparate} 
\begin{split}
\partial_t{\nabla^2\delta\chi} &= \frac{\Omega_{ci}}{B}\nabla^2\delta\phi + \upsilon^2_{t_i}\nabla^2\delta{n}+\frac{1}{2}\nabla^2|\bm\nabla\delta\chi|^2\\
&-\nu_{in}\nabla^2\delta\chi+\frac{4}{3}\frac{\upsilon^2_{t_i}}{\nu_{in}}\nabla^4\delta\chi\,,
\end{split}					
\end{align}
\begin{align}\label{eq:QuasiNeutralitySeparate}
\begin{split}
\partial_t\nabla^2\delta\phi &=-B\Omega_{ce}n_o^{-1}\bm{\nabla}\cdot(\delta{n}\bm{\nabla}\delta\chi)\\
&-B\Omega_{ce}n_o^{-1}\bm{\nabla}\cdot(n_o\bm{\nabla}\delta\chi)\\
&-\Omega_{ce}[\delta\phi,\ln{n_o}]-\Omega_{ce}[\phi_o,\delta{n}]-\Omega_{ce}[\delta\phi,\delta{n}]\\
&-\frac{1}{B}[\phi_o,\nabla^2\delta\phi]-\frac{1}{B}[\delta\phi,\nabla^2\delta\phi]\\
&+\frac{T_e\nu_{en}}{e}\nabla^2 \ln{n}+\frac{T_e\nu_{en}}{e}\bm{\nabla}\ln{n}\cdot\bm{\nabla}\ln{n}\\
&-\nu_{en}\nabla^2\phi-\nu_{en}\bm{\nabla}\ln{n}\cdot\bm{\nabla}\phi\,,\\
\end{split}					
\end{align}
where the gradients of the background density and electric potential are only defined in the vertical direction (\textit{i.e.,} $\partial_y{\ln n_o} = 0$ and $\partial_y\phi_o = 0$). 

To study the system's Hamiltonian we need to keep only the terms that are not injecting energy into the system (sources) or remove energy out of it (sinks). This is achieved by dropping any term that contains the gradient of the background density ($L_n^{-1} = \partial_z\ln n_o$) and/or background electric potential ($\upsilon_E = -B_o^{-1}\partial_z\phi_o$) which are considered energy sources in the dynamical system, and any viscosity term that contains a collision frequency of the electrons and/or ions with the background neutrals. This produces the following set of equations of motion:
\begin{align}\label{eq:IonContinuityHamiltonian}
\begin{split}
\partial_t{\delta{n}} = \bm\nabla\cdot(\delta{n}\bm\nabla{\delta\chi})\,,
\end{split}					
\end{align}
\begin{align}\label{eq:IonMomentumHamiltonian}
\begin{split}
\partial_t{\nabla^2\delta\chi} &= \frac{\Omega_{ci}}{B}\nabla^2\delta\phi + \upsilon^2_{t_i}\nabla^2\delta{n}+\frac{1}{2}\nabla^2|\bm\nabla\delta\chi|^2\,,
\end{split}					
\end{align}
\begin{align}\label{eq:QuasineutralityHamiltonian}
\begin{split}
\partial_t\nabla^2\delta\phi &=-B\Omega_{ce}n_o^{-1}\bm{\nabla}\cdot(\delta{n}\bm{\nabla}\delta\chi)\\
                             &-\Omega_{ce}[\delta\phi,\delta{n}]-\frac{1}{B}[\delta\phi,\nabla^2\delta\phi]\,.
\end{split}					
\end{align}
The set of \eqseqref{IonContinuityHamiltonian}{QuasineutralityHamiltonian}
shows only the dynamics in the fluctuating quantities \{$\delta{n},\delta\phi,\delta\chi$\} without including any sources or sinks of energy.\\

For the three evolving fields ($\delta{n},\delta\phi,\delta\chi$) in the dynamical system of the Equatorial Electrojet we can expect three components of energy; the electron's kinetic energy due to the $\bm{\delta{E}\times{B}}$ drifts, the ion's kinetic energy, and the internal thermal energy of both species. To check the way the fluctuating density is represented in the energy equation we use unknown functions of the density fluctuation ($\delta{n}$) and check the condition of zero rate of change of the system's Hamiltonian. So, we may propose a Hamiltonian for the system to be as follows:\\
\begin{eqnarray}
\label{eq:proposedH}
H &=& \int d^2\!x \bigg(
                  \frac{m_en_o}{2B^2}|\bm{\nabla}\delta\phi|^2 
                  \\
&& \hspace{1cm} +f(\delta{n})\frac{m_i}{2}|\bm{\nabla}\delta\chi|^2+m_i\upsilon^2_{t_i}g(\delta{n})
            \bigg)\,,	
            \nonumber			
\end{eqnarray}
where $f(\delta{n})$ and $g(\delta{n})$ are functions of the density fluctuation.\\

Solving for $f(\delta{n})$ and $g(\delta{n})$ that give zero rate of change of the Hamiltonian we get $f(\delta{n}) = \delta{n}$ and $g(\delta{n}) = \delta{n}^2/2$. Therefore, the energy equation can be written in terms of the three evolving field as:
\begin{eqnarray}
\label{eq:SysEnergy}
H &=& \int d^2\!x \bigg(
                  \frac{n_om_e}{2B^2}|\bm{\nabla}\delta\phi|^2
                   \\
                  && \hspace{1cm}
                  +\frac{m_i}{2}\delta{n}|\bm{\nabla}\delta\chi|^2+\frac{1}{2}m_i\upsilon^2_{t_i}\delta{n}^2)
            \bigg)\,.
            \nonumber
\end{eqnarray}

Most of the energy is found in the ion's kinetic and internal (thermal) energy parts of the total energy.\cite{hassan2015plasma,hassan2016multiscale} This is because the ion mass, which is found in the middle and last terms in equation(\ref{eq:SysEnergy}), is much larger than the electron mass, found in the electron's kinetic part of the total energy.\cite{hassan2016multiscale}

In addition, the rate of energy transfer in equation(\ref{eq:SysEnergy}), which manifests the rate at which the energy is injected, dissipated, and coupled between the evolving fields in the dynamical system, have the following form\cite{hassan2015plasma,hassan2016multiscale}:
\begin{eqnarray}\label{eq:EnergyComponents}
S_{\phi} &=&  \int\! d^2\!x\,  n_e T_e \upsilon_E\, \delta\tilde{\phi}\, \partial_y\delta\tilde{n}\,,\\
D_{\phi} &=&   \int\! d^2\!x\,  n_eT_e\rho_e^2\nu_{en}\,  \delta\tilde{\phi}\, \nabla^2\delta\tilde{\phi}\,,\\
C_{\phi\chi} &=&    \int\! d^2\!x\,  n_eT_e\, \delta\tilde{\phi}\nabla^2\delta\tilde{\chi}\,,\\
D^{(1)}_{\chi} &=&  - \int\! d^2\!x\, \frac{4}{3}\frac{n_iT_i}{\nu_{in}} \, \nabla^2\delta\tilde{\chi} \nabla^2\delta\tilde{\chi}\,,\\
D^{(2)}_{\chi} &=&    \int\! d^2\!x\,  n_im_i\nu_{in}\, \delta\tilde{\chi}\nabla^2\delta\tilde{\chi}\,. 
\end{eqnarray}
\begin{figure}
 \centering
 \begin{tabular}{c}
 \includegraphics[width=0.5\textwidth,height=0.4\textwidth]{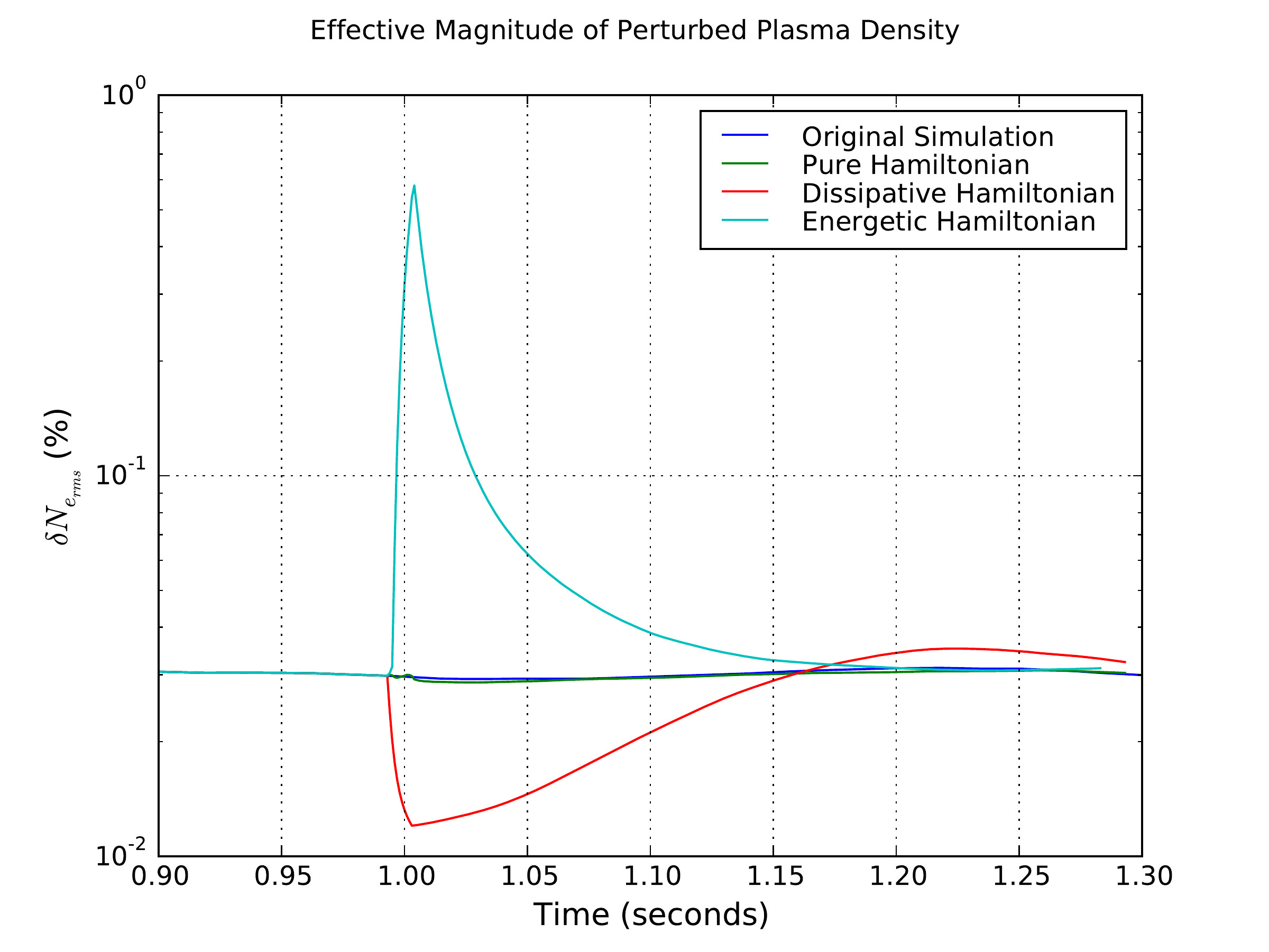}
 \end{tabular}
 \caption{\small A comparison between the physical role that different terms in Eq.~(\ref{eq:EnergyComponents}), in addition to the system Hamiltonian term, plays independently from the other terms during four different modes of simulation.} 
 \label{fig:EnergyTransfer}
\end{figure}
Equation (\ref{eq:EnergyComponents}) shows that the electron kinetic energy is responsible for injecting energy ($S_{\phi}$) into the system via the spatial variation of the electron density in the horizontal direction ($\partial_y\delta\tilde{n}$). However, the energy is dissipated in the viscosity of the electrons ($D_{\phi}$) and ions ($D^{(1)}_{\chi}$ and $D^{(2)}_{\chi}$) as they collide with the background neutrals. The energy is transferred between the electric potential ($\phi$) and ion velocity potential ($\chi$) via the coupling term ($C_{\phi\chi}$) which ensures the conservation of number of charged particles in the dynamical system. The dynamical  equation of the internal energy does not contribute in energy injection or dissipation.  It does, however, show strong coupling with the ions equation of motion.\cite{hassan2015plasma,hassan2016multiscale}

In Fig.~\ref{fig:EnergyTransfer} we studied the status of the dynamical system during the saturation region of the simulation by turning ON/OFF the energy injection and/or dissipation and comparing the root-mean-square values of the perturbed electron density. The blue line, which is barely seen as it is buried under the other lines, shows the status of the dynamical system when it includes all the physics such as the energy injection, dissipation, and the Hamiltonian. Because the system fulfills the fundamental law of energy conservation, that blue line is almost overlapped with the green line which represents the status of the dynamical system when simulating the system Hamiltonian. When the energy injection terms are turned off, the system dissipates its stored energy as the ions and electrons collide with background neutrals. After 10 ms we turned the energy injection terms on and we see the red line restore its original status. However, when the energy dissipation terms are turned off, the rate of growth of the perturbed electron density increases dramatically over the 10 ms. Once the energy dissipation terms are turned back on, the amplitude of the fluctuating density decreases to reach its original status as we can see in the cyan line. It can be easily noticed here that the rate of increase in energy when keeping the energy injectors but turning off the energy dissipation terms is much larger than the rate of decrease in energy when doing the opposite. This shows that the dynamical system responses faster to the energy injected compared to energy dissipation. Moreover, The restoring rate of dynamical system to its original state is much slower than the excitation rate when turning off the energy injection or dissipation terms. This is similar to Newton's law of cooling, where the rate of cooling the system is rapid at the beginning and slowing down as we go close to the room temperature which is pronounced clearly in the case of keeping the energy injectors and turning off/on the energy dissipation.

\section{Noncanonical Hamiltonian Structure}\label{sec:NonCanonicalHamiltonian}
\subsection{The Lie-Poisson bracket}\label{ssec:HamiltonianBracket}
Having a conserved energy is a necessary but not sufficient condition that would enable us to claim that the Equatorial Electrojet model of \eqseqref{IonContinuityHamiltonian}{QuasineutralityHamiltonian} is Hamiltonian. For this to happen, we need to find a Lie-Poisson bracket that would be antisymmetric, satisfy the Jacobi identity and reproduce the equations of motion. With the last statement, we mean that for an arbitrary field $\xi$, we could write it's time evolution as:
\begin{equation}
\frac{\partial \xi}{\partial t} = \{\xi,H\}\,,
\end{equation}
where $\{\cdot,\cdot\}$ is the Lie-Poisson bracket and $H$ the Hamiltonian of \eqeqref{SysEnergy}. The Lie-Poisson bracket will be a mathematical object of the form:
\begin{equation}
\{F,G\} = \int d^3\!x \frac{\delta F}{\delta \xi_i} \textbf{J}^{ij}\frac{\delta G}{\delta \xi_j}\,,
\end{equation}
where \textbf{J} is the so-called co-symplectic operator, $F$, $G$ are functionals and $\xi_i$, $\xi_j$ are dynamical fields of the system. 

To start building this bracket, we first need to calculate the functional derivatives of the Hamiltonian with-respect-to each evolving field, $\displaystyle\frac{\delta H}{\delta \xi_i}$. These functional derivatives can be read-off from the total variation of the Hamiltonian according to:
\begin{equation}
\delta H=\int d^2\!x \frac{\delta H}{\delta\xi_i} {\delta\xi_i}\,.
\end{equation}
After performing the variation on $H$, the aforementioned functional derivatives are easily found to be:
\begin{align}\label{eq:fun_deriv_H}
\begin{split}
\frac{\delta H}{\delta \zeta} &= -\frac{m_en_o}{ B^2}\phi\,, \\
\frac{\delta H}{\delta\chi} &= -m_i\bm{\nabla}\cdot\left(n\bm{\nabla}\chi\right)\,, \\
\frac{\delta H}{\delta{n}} &= \frac{m_i|\bm{\nabla}\chi|^2}{2}+m_i\upsilon^2_{t_i}n\,,
\end{split}
\end{align}
\noindent where $\zeta = \nabla^2\phi$.\\

Equipped with the previous relations for the functional derivatives of $H$ and noting the fact that the functional derivative of a field with respect to itself is a delta function,$\displaystyle \frac{\delta n(x)}{\delta n(x')} = \delta(x-x')$, we can work out a form for the bracket that reproduces all the equations of motion:
\begin{flalign}\label{eq:3rdBracket}
\{F,G\} &= -\frac{1}{m_i} \int d^2x (F_nG_{\chi}-G_nF_{\chi}) \nonumber\\
        &- \frac{B\Omega_{ci}}{m_en_o} \int d^2x (F_{\chi}G_{\zeta}-G_{\chi}F_{\zeta}) \nonumber\\
        &+ \frac{B^2\Omega_{ce}}{m_en_o^2} \int d^2x n[F_{\zeta},G_{\zeta}] \nonumber\\
        &+ \frac{B}{m_en_o} \int d^2x \zeta[F_{\zeta},G_{\zeta}]\,.
\end{flalign}
where by $F_{\xi}$, we mean the functional derivative $\displaystyle \frac{\delta F}{\delta \xi}$.

As already mentioned, the above constitutes only a candidate bracket for our system. To verify that the system is of the non-Canonical Hamiltonian type, the bracket needs to satisfy the Jacobi identity:
\begin{align}\label{eq:JacobiIdentity}
\{\{F,G\},H\} + \{\{H,F\},G\} + \{\{G,H\},F\} = 0\,,
\end{align}
with $F$,$G$ and $H$ being arbitrary functionals.

Indeed, the bracket of \eqeqref{3rdBracket} satisfies the Jacobi identity and the proof can be found in the appendix(\ref{app:JacobiIdenity}). 
\section{Casimir Invariants}\label{sec:Casimirs}
Casimirs are functionals found in noncanonical Hamiltonian systems that commute with every other functional in that system. As such, they are conserved quantities and constitute geometrical constants of motion. Therefore, they can be computed using the relation:
\begin{equation}\label{eq:Casimir}
\{F,C\} = 0\,,
\end{equation}
where $C$ is the Casimir functional and $F$ can be any functional of the system.

Casimir invariants are the result of degeneracy of the co-symplectic operator \textbf{J}. The gradients of the Casimir functionals, span the kernel of \textbf{J}.[Morrison 1998].  

\subsection{Finding the Casimir invariants}\label{ssec:FindingCasimirs}
To calculate the Casimirs of the Hamiltonian model of the Equatorial Electrojet, we invoke \eqeqref{Casimir} using the bracket of \eqeqref{3rdBracket}:
\begin{align}
\{F,C\} &= -\frac{1}{m_i} \int d^2\!x' (F_nC_{\chi}-C_nF_{\chi}) \nonumber\\
        &- \frac{B\Omega_{ci}}{m_en_o} \int d^2\!x' (F_{\chi}C_{\zeta}-C_{\chi}F_{\zeta}) \nonumber\\
        &+ \frac{B}{m_en_o} \int d^2\!x' \zeta[F_{\zeta},C_{\zeta}] \nonumber\\
        &+ \frac{B^2\Omega_{ce}}{m_en_o^2} \int d^2\!x' n[F_{\zeta},C_{\zeta}] = 0\,. \nonumber\\
\end{align}

Now, we factor out the different variations of $F$:
\begin{align}
\{F,C\} &= \int d^2\!x' F_n\left(-\frac{1}{m_i}C_{\chi}\right) \nonumber\\
        &+ \int d^2\!x' F_{\chi} \left(\frac{1}{m_i}C_n-\frac{B\Omega_{ci}}{m_en_o}C_{\zeta}\right) \nonumber\\
        &+ \int d^2\!x' F_{\zeta}\left(\frac{B^2\Omega_{ce}}{m_en_o^2}[n,C_{\zeta}]+\frac{B}{m_en_o}[\zeta,C_{\zeta}]\right) = 0\,.
\end{align}

Because the variations on $F$ are independent of each other, we can deduce conditions for the vanishing of $\{F,C\}$. These are written as follows:

\begin{flalign}\label{eq:cas1}
C_{\chi} = 0
\end{flalign}

\begin{flalign}\label{eq:cas2}
C_n = \frac{B\Omega_{ce}}{n_o} C_{\zeta}
\end{flalign}

\begin{flalign}\label{eq:cas3}
\left[B\Omega_{ce}n+n_o\zeta, C_{\zeta}\right]=0
\end{flalign}

From \eqeqref{cas1} we can surmize that the Casimirs will not be a function of $\chi$, whereas \eqseqref{cas2}{cas3} are equivalent and force us to the conclusion that the system has an infinite family of Casimirs that are of the form:

\begin{flalign}\label{eq:CasimirIntegral}
C = \int d^2\!x f(B\Omega_{ce}n+n_o\zeta)\,.
\end{flalign}

\subsection{Normal Fields -- Reformulation in new variables}\label{ssec:newVariables}
The form of the Casimir Invariants obtained in \ssecref{FindingCasimirs} suggests the introduction of a new variable, $Q = B\Omega_{ce}n+n_o\zeta$. Indeed, if we use \eqseqref{IonContinuityHamiltonian}{QuasineutralityHamiltonian} to calculate the time evolution of the quantity $Q$, we arrive at:

\begin{flalign}\label{eq:NewVariable}
\frac{\partial Q}{\partial t} +[F_{E{\times}B},Q] = 0\,,
\end{flalign}
where $\displaystyle F_{E{\times}B}={\phi}/{B}$. 

From \eqeqref{NewVariable} we observe that the so-called ``normal field'' $Q$ is a Lagrangian invariant of the system since it is only advected by the stream function $F_{E{\times}B}$. 

Now, we can use the new varible $Q$ to re-express the bracket of \eqeqref{3rdBracket}. To make this change of variables from $F[n,\zeta,\chi]$ to $\bar{F}[n,Q,\chi]$, we first need to rewrite all the functional derivatives using the following relation:

\begin{flalign}
\int d^2\!x \left(\frac{\delta F}{\delta n}\delta n + \frac{\delta F}{\delta \zeta}\delta \zeta + \frac{\delta F}{\delta \chi}\delta \chi\right) &= \nonumber\\
\int d^2\!x \left(\frac{\delta \bar{F}}{\delta n}\delta n + \frac{\delta \bar{F}}{\delta Q}\delta Q + \frac{\delta \bar{F}}{\delta \chi}\delta \chi\right)\,.& \nonumber
\end{flalign}

Using the fact that $\delta Q = {B\Omega_{ce}\delta n}+{n_o\delta\zeta}$ and comparing both sides of the previous equation, we get:

\begin{flalign}
F_n &= \bar{F}_n+B\Omega_{ce}\bar{F}_Q \,,\nonumber\\
F_{\zeta} &= n_o \bar{F}_Q\,, \\
F_{\chi} &= \bar{F}_{\chi}\,. \nonumber
\end{flalign}

Thus, we can rewrite the bracket of \eqeqref{3rdBracket} in terms of the new variable as follows:

\begin{flalign}
\{F,G\} &= -\frac{1}{m_i} \int d^2\!x \left(F_nG_{\chi}-G_nF_{\chi}\right) \nonumber\\
        &+ \frac{B}{m_e}\int d^2\!x Q[F_Q,G_Q]\,. \nonumber
\end{flalign}
where we have dropped the overbars.

Moreover, the dynamical equations of the Equatorial Electrojet model can be rewritten as:

\begin{flalign}
\partial_t{n} &= \bm\nabla \cdot (n \bm\nabla{\chi})\,, \\
\partial_tQ &= [Q,F_{E{\times}B}]\,, \label{eq:newQuasineutrality}\\
\partial_t\nabla^2\chi &= \frac{\Omega_{ci}}{n_oB}Q-\frac{\Omega_{ce}B}{n_o}n \nonumber\\
                       &+\upsilon_{t_i}\nabla^2\ln n+\frac{1}{2}\nabla^2|\bm{\nabla}\chi|^2\,.
\end{flalign}

The new form of \eqeqref{newQuasineutrality} which is based on the quasineutrality condition of the plasma, shows the dependence of plasma dynamics (to drive the equatorial electrojet instabilities) on the density gradient and the $\bm{{E}\times{B}}$ drifts which are representing the energy sources in the dynamic systems. However, the other two dynamical equations, that control the coupling and dissipation in the system, do not show any change in their structure. This emphasizes  the essential role that the electron density and electric potential play in evolving the  dynamical system and generating the active turbulent structures in the equatorial electrojet, which can  be seen in the radar backscattered echoes and rocket observations.

\section{Simulation results}\label{ssec:simulationResults}
A validation of the unified fluid model described in Sec.~\ref{sec:PlasmaDynamics} was  carried out by comparing the simulation results to the radar observations and sounding rocket measurements under different solar and geophysical conditions.\cite{hassan2015multiscale,hassan2016multiscale} The Gradient-drift and Farley-Buneman instabilities are found to be excited simultaneously in the equatorial electrojet. Whereas the observation of the Farley-Buneman instability depends on the availability of a cross-field drift that exceeds the ion-acoustic speed, the Gradient-drift instability is found in the presence of a sharp positive density-gradient in the ionosphere. Therefore, both instabilities can be observed simultaneously when the condition of each is realized in the electrojet. However the strong backscattered echoes of the Farley-Buneman instability sometimes block the observation of the Gradient-drift instability and make it invisible. Therefore, the presence of the short and long plasma waves due to the Farley-Buneman and Gradient-drift instabilities is not related to their observance in the backscattering echoes.

In Ref.~\onlinecite{hassan2015multiscale} the authors presented several linear and nonlinear simulations results that were able to distinguish between the different types of plasma waves resulting from different instability mechanisms, and how the small structures (formed due to the breakup of the large structures into small ones) fill the simulation box in such a way that explains the invisibility of the large-scale structures from the radar echoes in the presence of small-scale structures.

In Ref.~\onlinecite{hassan2016multiscale} the authors presented the energy distribution over structures of different scale lengths and the cascading of energy from the large-scale structures into small-scale ones,  which emphasizes  the prevalence of  echoes that backscatter  from small structures in the radar observation.

In this subsection we present new  linear and nonlinear results for the phase velocity, the phase relationship between the plasma density and the components of the perturbed electric field, and the effect of the free energy sources in the dynamical  system on the effective magnitude of the electric field in the electrojet. These simulation results further validate the ability of the unified fluid model to simulate the plasma dynamics and instabilities in the equatorial electrojet.

\subsection{Growth-rate and Phase Velocity}
The linear calculations  presented in Refs.~\onlinecite{hassan2015multiscale,hassan2015plasma} show a strong dependency of the growth-rate on the local values of the ionospheric parameters such as the density-gradient scale-length and background electric field. The vertical growth-rate profile divides the entire electrojet into three regions.  The bottom one (90--103 km) is dominated by the gradient-drift instability when the cross-field drift speed is smaller than the ion-acoustic speed, and the unstable plasma waves result from the Farley-Buneman instability can not be excited.\cite{pfaff1987electricA,pfaff1987electricB}  In the top region of the electrojet (108--120 km) the density profile is inverted due to the presence of the E-region nose as a result of the decrease in the density, and this inhibits the generation of  unstable large-scale plasma waves that are excited by  the presence of  a sharp positive density-gradient.\cite{farley1973instabilities,pfaff1987electricA,pfaff1987electricB}  Therefore, the top region of the electrojet is dominated by the small-scale structures results from the excitation of Farley-Buneman instability as the cross-field drift speed is always larger than the ion-acoustic speed.\cite{hassan2015multiscale}  The core of the equatorial electrojet (103--108 km) is found to be very rich with unstable waves of all wavelengths result from the coupling between the gradient-drift and Farley-Buneman instabilities, and the electrojet current has its peak value in that region where the electrical conductivity is found to have its maximum value in the ionosphere at this region of the electrojet.\cite{baker1953electric,farley2009equatorial} 
\begin{figure}
 \centering
 \begin{tabular}{c}
 \includegraphics[width=0.4\textwidth,height=0.3\textwidth]{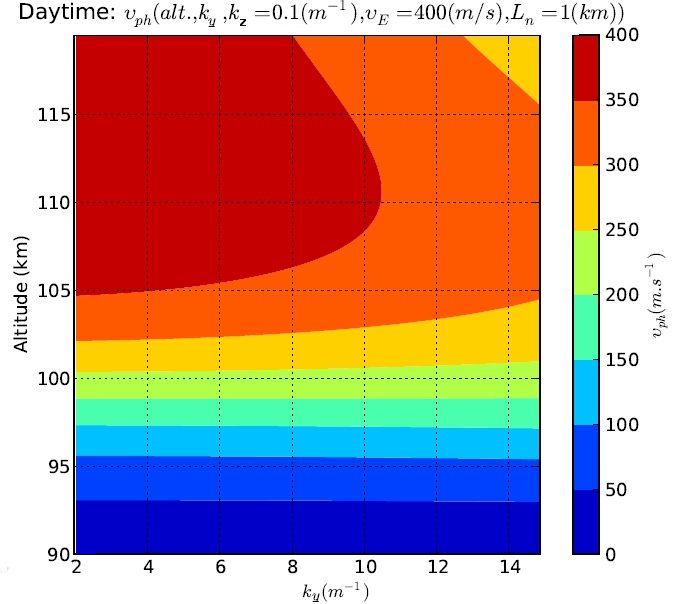}
 \end{tabular}
 \caption{\small A vertical profile of the linear phase velocity as a function of altitude that has a maximum value equal to the background cross-field drift speed ($\upsilon_E$). The linear phase velocity profile is divided into three regions; gradient-drift dominant region between 90--103 km, Farely-Buneman dominant region above 110 km, and a coupling region between both instabilities between 103 and 110 km.} 
 \label{fig:PhaseVelocity}
\end{figure}

A similar result can be seen in Fig.~\ref{fig:PhaseVelocity},  where the phase velocity is calculated at different altitudes using nonlocal magnitudes of the ionospheric background.\citep{hassan2013nonlocal,hassan2014equatorial}  We still can see the three distinct regions of the equatorial electrojet with the exchange dominance and coupling between unstable waves of different scale-sizes that are generated as a result of the gradient-drift and Farley-Buneman instabilities. The maximum phase velocity is found at the core of the electrojet around 105 km has the same magnitude of the $\bm{E}\times \bm{B}$ drift velocity (400 m/s). However, the radar observations and rocket measurements found the maximum drift value of the electrojet to be equal to the local value of the ion-acoustic speed ($\simeq$ 320 m/s at 105 km). Therefore, the linear results fail to explain the decrease in the electrojet drift speed below the cross-field drift speed.

\subsubsection{Density and Electric Field Phase Relationship}
\begin{figure}
 \centering
 \begin{tabular}{c}
 \includegraphics[width=0.5\textwidth,height=0.4\textwidth]{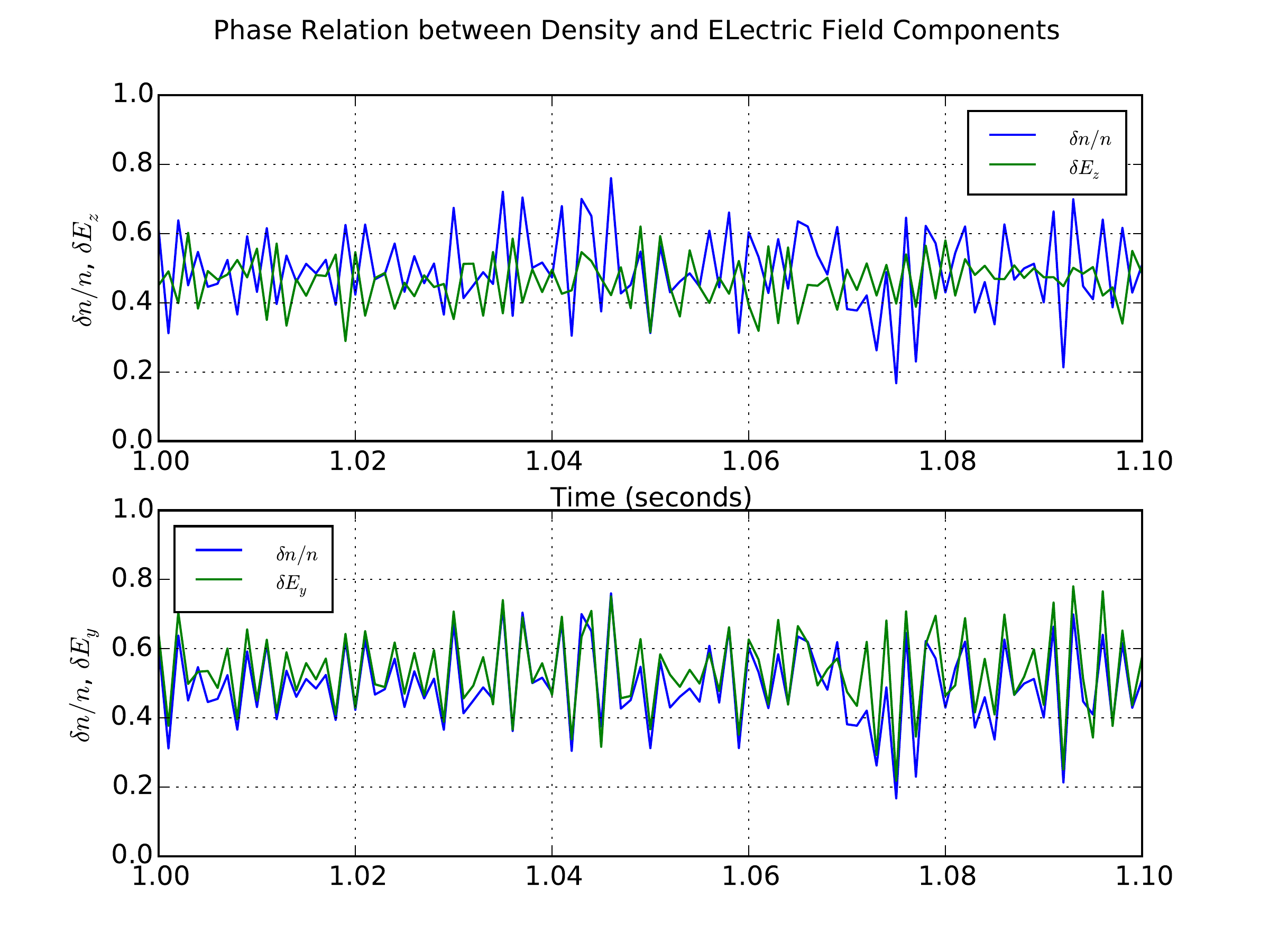} \end{tabular}
 \caption{\small A comparison of the phase relationship between the normalized quantities of the perturbed density and the perturbed component of the electric field in the horizontal and vertical directions. There is an in-phase relationship between the density and the horizontal component of the electric field, however the phase relationship between the density and the vertical component of the electric field is mostly out-of-phase.} 
 \label{fig:DEFComp}
\end{figure}

The phase differences between the fluctuations in the plasma density and electric field components are important to understand the excitation of the plasma instabilities in the equatorial electrojet region. In figure(\ref{fig:DEFComp}) we see the phase relationship between the perturbed density (blue-line) and the vertical (top-panel) and horizontal (bottom-panel) perturbed electric field components (green-line) at 105 km altitude with all of these quantities normalized. The in-phase relationship between the plasma density and the horizontal component of the electric field emphasize on the excitation of the gradient-drift instability in the horizontal direction and the generation of long-wavelength plasma structures in the electrojet as measured by the sounding rocket during the CONDOR campaign \cite{pfaff1987electricA}. However, the out-of-phase relationship between the plasma density and the vertical component of the electric field explains the growing of the unstable waves in the vertical direction and the generation of the small-scale structures due to the Farley-Buneman instability when the cross-field drift exceeds the ion-acoustic speed \cite{kelley2009earth}. Beside showing a good agreement with the rocket measurements, these simulation results emphasize on the validity of using the unified fluid model in studying the energy cascading in the equatorial electrojet between plasma irregularities of long and short scale sizes that are excited in the horizontal and vertical directions, respectively.

\subsection{Energy Sources and Electric Fields}
The vertical gradients of the background plasma density and electric potential are considered the source of free energy coming into the system throughout its boundaries.\cite{hassan2016multiscale}  Both of these gradients ($\partial_zn_o$ and $\partial_z\phi_o$) give rise to two electron drifts in the westward directions of different speeds and scales.\cite{fejer1980ionospheric} The density-gradient gives rise to a slow drift of large-scale irregularities, however the potential-gradient of a proper magnitude gives rise to an ultrasonic drift of meter-scale structures.\cite{farley1973instabilities,farley2009equatorial} The amount of energy injected into the system depends on the sharpness of the gradient of these ionospheric background quantities.

\begin{figure}
 \centering
 \begin{tabular}{c}
 \includegraphics[width=0.5\textwidth,height=0.4\textwidth]{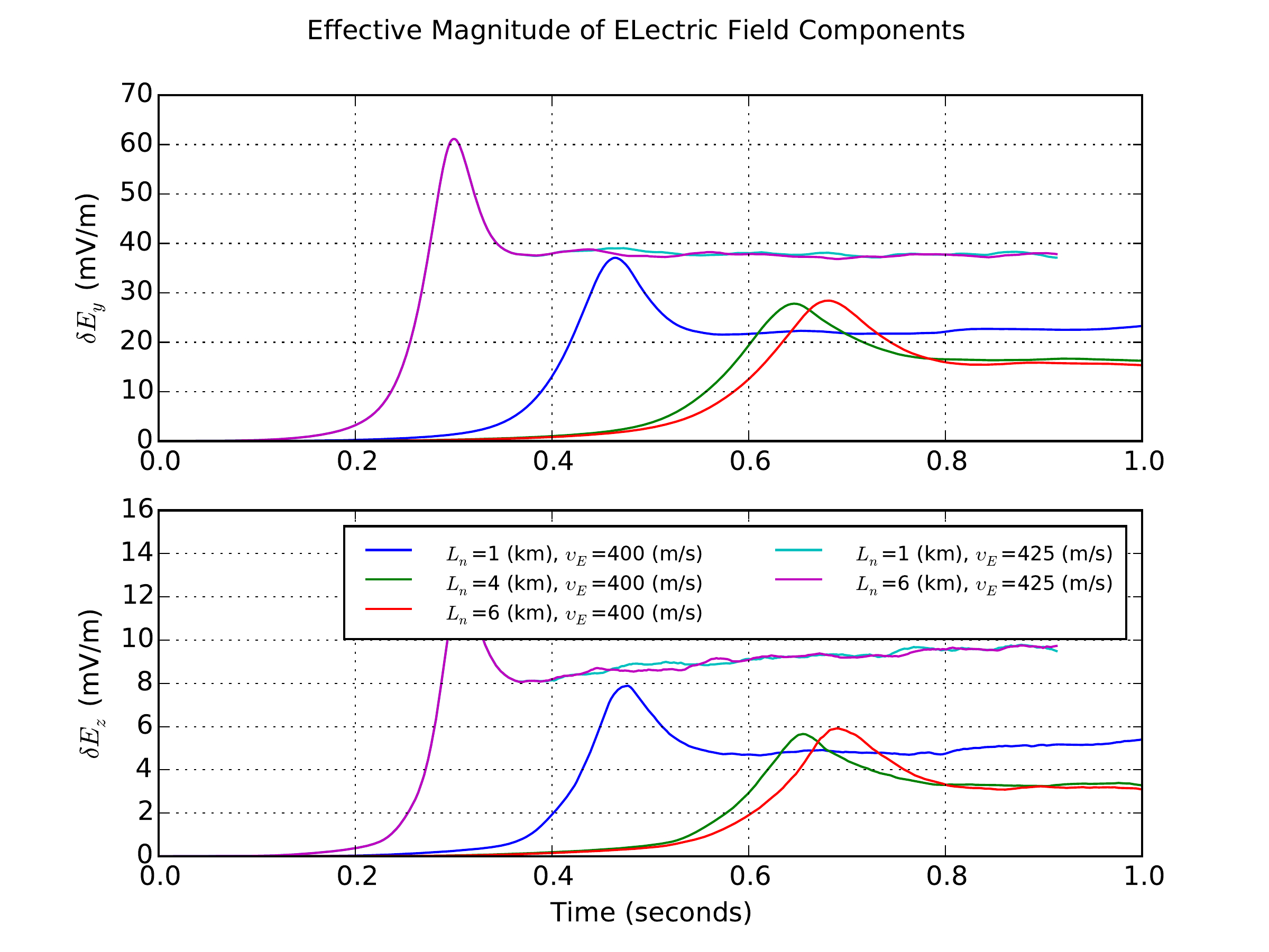} \end{tabular}
 \caption{\small A comparison between the influence of the energy injected into the system from the boundary in the form of cross-field drift-velocity ($\upsilon_E$) and positive density-gradient scale-length ($L_n$) on the perturbed components of the electric field in the horizontal (upper-panel) and vertical (lower-panel) directions. The increase in the horizontal component of the $\bm{E}\times \bm{B}$ drift-velocity has a larger effect on the saturation level of the perturbed electric field compared to a similar increase in the scale-length of the density-gradient. The effect of increasing ($L_n$) is found to disappear for very large values of ($\upsilon_E$).}
 \label{fig:EFComp}
\end{figure}

The results of multiple simulations of the horizontal ($\delta{E_y}$) and vertical ($\delta{E_z}$) components of the perturbed electric field at the core of the equatorial electrojet at different gradients of the plasma density and the electric potential are shown in Fig.~\ref{fig:EFComp}. It can be noticed that all simulations run through three distinct phases; the linearly dominant growing, the transitional, and the saturation phase, and the rate of growing of the unstable modes in the system and consequently the saturation level of electric field components are highly dependent on the magnitude of the density-gradient scale-length ($L_n = n_o\partial^{-1}_zn_o$) and the cross-field drift ($\upsilon_E = - B^{-1}\partial_z\phi_o$). A close look at the simulation results in Fig.~\ref{fig:EFComp} tells us that a small increase in the $\bm{E}\times \bm{B}$ has a larger impact on the rate of growing of the unstable plasma waves and magnitude of the perturbed electric field at the saturation phase compared to a similar or larger difference in the density-gradient scale length ($L_n$). We can see that a change in $L_n$ from 6 to 4 km does not affect the effective value of the perturbed horizontal and vertical electric field components in the saturation phase which set on at $E_y$=18 (mV/m) and $E_z$=3.5 (mV/m), respectively. However, a small change in $\upsilon_E$ from 400 to 425 (m/s) pumps enough energy into the dynamical  system to double the magnitude of the electric field components at the saturation phase.

Therefore, we can conclude that the cross-field drift speed pumps a larger amount of energy into the dynamical  system compared to the density-gradient drift. Because Farley-Buneman instability depends mainly on the magnitude of the cross-field drift, the small-scale structures will have more energy compared to the large-scale ones excited in the system as a result of the gradient-drift instability. This elucidates the dominance of the spectrum of the short-scale structures in the radar echoes of the electrojet irregularities along with the absence of the spectrum of the large-scale structure. The spectrum of the unstable waves generated by the gradient-drift instability can only be seen when the cross-field drift is smaller than the ion-acoustic speed, i.e. during the absence of the small-scale structures in the equatorial electrojet.\cite{fejer1975vertical,kelley2009earth} 

\section{Summary and Conclusions}\label{sec:conclusions}
The plasma dynamics in the unified fluid model for the Equatorial Electrojet instabilities which is proposed in Ref.~\onlinecite{hassan2015multiscale} and validated against the rocket measurements and radar observations is discussed. In addition, more linear and nonlinear results are presented to provide another validation test of the model to simulate the equatorial electrojet instabilities and showed a good agreement with the available observations.

In the simulation results, the linear phase velocity fails to explain the bounding of the electrojet speed to the local magnitude of the ion-acoustic speed, however the relative speed between the electrons and ions during the saturation phase of the nonlinear simulation is found to be so close to the ion-acoustic speed calculated at the core of the electrojet. Also, the in-phase relationship between the plasma density and the horizontal component of the perturbed electric field verifies the presence of plasma structures of large wavelength in the horizontal direction. However, the out-of-phase relationship between the vertical component of the electric field and the plasma density supports the presence of the small-scale unstable waves in the vertical direction which is excited as a result of the Farley-Buneman instability.

The effect of the variations in the available free energies in the dynamical system on the effective magnitude of the perturbed horizontal and vertical components of the electric field is examined based on multiple simulation results. The cross-field drift is found to provide large amount of free energy in the system compared to the density-gradient drift. In addition, any small change in the cross-field drift manifests itself as a large leap in the magnitude of the electric field components, but this is not the case with the density-gradient drift which shows only a big difference during the growing phase of the simulation but not in the steady-state phase. Moreover, for large values of the cross-field drift the change in the density-gradient scale-length does not show any effect in the magnitude of the electric field components over all phases of the simulations. This explains the absence of the spectrum of the large-scale structures from the radar echoes during the excitation of the Farley-Buneman instability.

The model proposed here for the Equatorial Electrojet was found to be a noncanonical Hamiltonian system and a Lie-Poisson bracket for it, that satisfies the Jacobi identity, has been given. Moreover, an infinite family of Casimir Invariants for the system has been identified and used to reformulate it in a way which brings to light a Lagrangian invariant. 

The unified fluid model captures the characteristics of plasma instabilities (Gradient-drift and Farley-Buneman) in the equatorial electrojet based on the properties of the ionospheric plasma in the E-region, such as plasma density, electric field, and temperature in a 2-D geometry. The applicability of the unified model in the high-latitude electrojet is restricted, however,  because  the incorporation of the electron and ion heating into the model would require  extending the system geometry into the third dimension (along the magnetic field lines), and adding the proper physics that describes the required sources of energy dissipation that help to stabilize  the generated plasma instabilities. Therefore, the unified fluid model in its 2-D geometry works well in the equatorial electrojet, but fails to model and capture  characteristics of the plasma instabilities in the high-latitude electrojet.

\appendix
\section{Dynamical System Brackets}\label{app:functionalDerivative}
\noindent Here, we show the procedure we followed to arrive at the bracket of \eqeqref{3rdBracket}. First, we start by expressing the dynamical equations in integral form. We will only display the example of \eqeqref{QuasineutralityHamiltonian}:
\begin{flalign}
\partial_t\zeta &= \frac{\Omega_{ce}}{n_o} \int d^2\!x' \delta(x'-x)[n,\phi] \nonumber\\
                &-\frac{1}{B} \int d^2\!x' \delta(x'-x) [\phi,\zeta] \nonumber\\
                &-\frac{B\Omega_{ce}}{n_o} \int d^2\!x' \delta(x'-x)\bm{\nabla}\cdot(n\bm{\nabla}\chi)\,.\nonumber\\
\end{flalign}
Next, we perform an integration by parts to place the delta functions inside the Poisson brackets and we invoke the relations for functional derivatives of the Hamiltonian of \eqeqref{fun_deriv_H}:
\begin{flalign}
\partial_t\zeta &=-\frac{\Omega_{ce}}{n_o} \int d^2\!x' n[\delta(x'-x),\phi] \nonumber\\
                & -\frac{1}{B} \int d^2\!x' \zeta[\delta(x'-x),\phi] \nonumber\\
                & +\frac{B\Omega_{ce}}{m_in_o} \int d^2\!x' \left(\frac{\delta\zeta}{\delta\zeta}\frac{\delta{H}}{\delta\chi}-\frac{\delta{H}}{\delta\zeta}\frac{\delta{\zeta}}{\delta\chi}\right)\,. \nonumber
\end{flalign}

Finally, we use the fact that the functional derivative of a field with respect to itself gives a delta function, to rewrite the above as:
\begin{flalign}
\partial_t\zeta &= \frac{B^2\Omega_{ce}}{m_en_o^2} \int d^2\!x' n\left[\frac{\delta\zeta}{\delta\zeta},\frac{\delta{H}}{\delta\zeta}\right] \nonumber\\
                &+ \frac{B}{m_en_o} \int d^2\!x' \zeta\left[\frac{\delta\zeta}{\delta\zeta},\frac{\delta{H}}{\delta\zeta}\right] \nonumber\\
                &-\frac{B\Omega_{ci}}{m_en_o} \int d^2\!x' \left(\frac{\delta{\zeta}}{\delta\chi}\frac{\delta{H}}{\delta\zeta}-\frac{\delta{H}}{\delta\chi}\frac{\delta\zeta}{\delta\zeta}\right)\,. \label{zetaBracket}
\end{flalign}

Consequently, taking advantage of the fact that $\partial_t{\nabla^2\phi} = \{\nabla^2\phi,H\}$, we postulate that a suitable form for a bracket would be:
\begin{flalign}
\{F,G\} &= \frac{B^2\Omega_{ce}}{m_en_o^2} \int d^2\!x' n\left[F_\zeta,G_\zeta\right] \nonumber\\
                &+ \frac{B}{m_en_o} \int d^2\!x' \zeta\left[F_\zeta,G_\zeta\right] \nonumber\\
                &-\frac{B\Omega_{ci}}{m_en_o} \int d^2\!x' \left(F_\chi G_\zeta - F_\zeta G_\chi\right)\,. \label{zetaBracket}
\end{flalign}

In a similar fashion, we can work out the remaining terms of the bracket, performing the same procedure to the rest of the equations of motion.

\section{Jacobi Identity}\label{app:JacobiIdenity}
The functional derivatives of the brackets with-respect-to the evolving fields ($n,\phi,\chi$) are given by:
\begin{flalign}\label{eq:bracket_derivatives}
\begin{split}
\{A,B\}_n &= \frac{B^2\Omega_{ce}}{m_en_o}[A_{\zeta},B_{\zeta}]\,, \\
\{A,B\}_{\chi} &= 0 \,,\\
\{A,B\}_{\zeta} &= \frac{B}{m_en_o}[A_{\zeta},B_{\zeta}]\,.
\end{split}
\end{flalign}

We proceed with the calculation of the first term of the Jacobi identity:
\begin{align}
\begin{split}\nonumber
\{\{A,B\},C\} &=  -\frac{1}{m_i}\int d^2\!x \left(\{A,B\}_nC_{\chi}-C_n\{A,B\}_{\chi}\right) \\
  &~~~~ -\frac{B\Omega_{ci}}{m_en_o}\int d^2\!x \left(\{A,B\}_{\chi}C_{\zeta}-C_{\chi}\{A,B\}_{\zeta}\right)\\
  &~~~~ +\frac{B^2\Omega_{ce}}{m_en_o}\int d^2\!x\, n\left[\{A,B\}_{\zeta},C_{\zeta}\right] \\
  &~~~~ +\frac{B}{m_en_o}\int d^2\!x\, \zeta\left[\{A,B\}_{\zeta},C_{\zeta}\right]\,.
\end{split}
\end{align}

Substituting \eqeqref{bracket_derivatives} in the above, we find:
\begin{flalign}
\begin{split}\nonumber
\{\{A,B\},C\} &= -\frac{B^2\Omega_{ci}}{m_e^2n_o^2}\int d^2\!x [A_{\zeta},B_{\zeta}] C_{\chi} \\
              &  +\frac{B^2\Omega_{ci}}{m_e^2n_o^2}\int d^2\!x [A_{\zeta},B_{\zeta}] C_{\chi} \\
              &  +\frac{B^3\Omega_{ce}}{m_e^2n_o^2}\int d^2\!x \,n\left[[A_{\zeta},B_{\zeta}],C_{\zeta}\right] \\
              &  +\frac{B^2}{m_e^2n_o^2}\int d^2\!x \,\zeta\left[[A_{\zeta},B_{\zeta}],C_{\zeta}\right]\,.
\end{split}
\end{flalign}

Combining terms together, we arrive at:
\begin{align}\label{eq:1stJacobiTerm}
\{\{A,B\},C\} &=  \frac{B^3\Omega_{ce}}{m_e^2n_o^2}\int d^2\!x \,n\left[[A_{\zeta},B_{\zeta}],C_{\zeta}\right] \nonumber\\
              &  +\frac{B^2}{m_e^2n_o^2}\int d^2\!x \,\zeta\left[[A_{\zeta},B_{\zeta}],C_{\zeta}\right]\,.
\end{align}

Now, it is obvious to see that the bracket satisfies the Jacobi identity since the inner bracket, which is simply a Poisson bracket $([\cdot,\cdot])$ has this property.

\section*{Acknowledgment}

\noindent This work was  supported by U.S. Dept.\ of Energy  under contract \#DE-FG02-04ER-54742.  PJM would  like to acknowledge support from the Humboldt Foundation and the hospitality of the Numerical Plasma Physics Division of the IPP, Max Planck, Garching. 

\nocite{*}
\bibliography{EEJHamiltonian}

\end{document}